# SaSeVAL: A Safety/Security-Aware Approach for Validation of Safety-Critical Systems

1st Wolschke, Christian
*Fraunhofer IESE*
Kaiserslautern, Germany
christian.wolschke@iese.fraunhofer.de

2nd Sangchoolie, Behrooz
*RISE Research Institutes of Sweden*
Borås, Sweden
behrooz.sangchoolie@ri.se

3rd Simon, Jacob
*CEVT, China Euro Vehicle Technology*
Göteborg, Sweden
jacob.simon@cevt.se

4nd Marksteiner, Stefan
*AVL List GmbH*
Graz, Austria
stefan.marksteiner@avl.com

5th Braun, Tobias
*Fraunhofer IESE*
Kaiserslautern, Germany
tobias.braun@iese.fraunhofer.de

6th Hamazaryan, Hayk
*ZF Friedrichshafen AG*
Friedrichshafen, Germany
hayk.hamazaryan@zf.com

*Abstract*—Increasing communication and self-driving capabilities for road vehicles lead to threats imposed by attackers. Especially attacks leading to safety violations have to be identified to address them by appropriate measures. The impact of an attack depends on the threat exploited, potential countermeasures and the traffic situation. In order to identify such attacks and to use them for testing, we propose the systematic approach SaSeVAL for deriving attacks of autonomous vehicles.

SaSeVAL is based on threats identification and safety-security analysis. The impact of automotive use cases to attacks is considered. The threat identification considers the attack interface of vehicles and classifies threat scenarios according to threat types, which are then mapped to attack types. The safety-security analysis identifies the necessary requirements which have to be tested based on the architecture of the system under test. It determines which safety impact a security violation may have, and in which traffic situations the highest impact is expected. Finally, the results of threat identification and safety-security analysis are used to describe attacks.

The goal of SaSeVAL is to achieve safety validation of the vehicle w.r.t. security concerns. It traces safety goals to threats and to attacks explicitly. Hence, the coverage of safety concerns by security testing is assured. Two use cases of vehicle communication and autonomous driving are investigated to prove the applicability of the approach.

## I. INTRODUCTION

New interaction capabilities in-between vehicles (e.g. V2V communication, cooperative services), better sensor technologies, and more advanced sensor data processing (e.g. camera-based AI object recognition) lead to the demand to test the systems also against cybersecurity threats. The required sensors and controls significantly increase the attack surface of modern vehicle in terms of cyber attacks. Furthermore, due to the increasing influence of assistance systems regarding vehicle control and autonomy, the potential severity of incidents resulting from malfunction caused by such attacks increases. Since cybersecurity threats may have a direct impact on the safety of the overall system, it is of uttermost importance that all threats that could potentially lead to safety goal violations are systematically identified and removed.

In this paper, we present SaSeVAL, a Safety/Security-aware approach for VALidation of safety-critical systems which we applied to two automotive use-cases. It aims at specifying all potential attacks, which may have a safety impact. The attacks are supposed to be used for the validation of the system under test (SUT). The approach has been developed and evaluated in the context of automotive use cases, but it as generic that it could be also used in other safety-critical domains.

The appropriateness of the SUT reaction depends on the correct implementation of the security control in a given situation. Moreover, the SUT must react within a time period so that safety impacts are avoided. The idea is inspired by ISO 26262 [6] in which *fault tolerant time intervals* (FTTIs) are assigned to safety goals, so that the counter measures of the SUT have a maximum time span to react and mitigate the imminent hazardous event. As it could be difficult to determine appropriate reaction times in practice, we decided to specify the situations in which the SUT could be attacked.

To this end, we address the following research questions:
- **RQ1**: How can we ensure completeness in safety-security co-engineered automotive validation testing?
- **RQ2**: How can we judge the severity of threats to reduce the test space?
- **RQ3**: How can we specify attacks so that testers can correctly reproduce them during cybersecurity testing?

In order to achieve completeness of security tests, it is necessary to identify all threats that could adversely affect the SUT and to evaluate the security controls of the system in all critical situations and with respect to the threats identified. In this paper, we present a systematic approach to identify the threats and to store them in a *threat library*. The identification of critical test situations requires a systematic risk assessment. We enrich the notion of the Threat Assessment

This work has received funding from ECSEL Joint Undertaking (JU) under grant agreement No 783119. The JU receives support from the European Union Horizon 2020 research and innovation program and Netherlands, Austria, Belgium, Czech Republic, Germany, Spain, Finland, France, Hungary, Italy, Portugal, Romania, Sweden, United Kingdom, Tunisia.

and Remediation Analysis (TARA) to create an explicit link to safety analysis of ISO 26262 (see §II-B and §II-C).

The paper is structure as follows. In §II, some background connected to the security engineering approaches for vehicles as well as the TARA approach are presented. The SaSeVAL approach is then detailed in §III. We then applied SaSeVAL to two automotive related use cases and share the experiences obtained in §IV. Finally, conclusions are drawn for the applicability to other projects and with respect to further improvements, which are presented in §V.

## II. BACKGROUND

In this section, we first present some backgrounds related to security engineering for vehicles. We then present TARA [9] as well as HARA of ISO 26262 as SaSeVAL uses them for identification of threats and analysis of safety, respectively.

### A. Security Engineering for Vehicles

Security engineering for vehicles is standardized in SAE J3061 - *Cybersecurity Guidebook for Cyber-Physical Vehicle Systems* [13]. Additionally ISO/SAE 21434 *Road vehicles — Cybersecurity engineering* [4] is under development with the goal to establish a standard for the Cybersecurity Management System of vehicles. The approach proposed in this paper (see §III) is suitable to be used at the concept phase of the development process in which threat analysis and risk assessment (TARA) are carried out and cybersecurity concepts are created. Our attack descriptions (see §III-C) are suited for formulating attacks in the early stages of the development process as well as for testing a system against these when performing validation and verification activities.

The risk assessment for cybersecurity threats investigated in this paper is based on the notion that the risk depends on the asset, the threat, and the vulnerability [7] [8] [5]. The risk assessment is based on the ISO 26262 standard [6] which addresses functional safety in road vehicles. ISO 26262 recommends to rate risks of failures according to the severity, exposure, and the controllability.

Pekaric et al. [11] have conducted a literature survey to investigate the application of security testing techniques for automotive systems. They show that most of work conducted in the past focus on penetration testing or the application of model-based testing. While some papers address risk-based security testing, the authors of the literature survey identify a lack of safety-driven security testing. This kind of testing is in fact of utmost importance as the system under tests are safety-critical, meaning that failures in them could result in loss of lives or could result in environmental damages.

Security testing of vehicles is conducted differently compared to the security testing [12] of information systems. Attacks may occur at the in-vehicle network structure, the wireless access or the physical access to the system under test. Another difference in automotive attack testing are the potential attackers, which are identified as: vehicle owner/driver, evil mechanic, thief and remote attacker [12] [2].

### B. Threat Analysis and Risk Assessment (TARA)

Threat Analysis and Risk Assessment (TARA) facilitates identification of threats, evaluation and assignment of respective risks as well as analysis of potential benefits of security controls for reducing the probability of successful attacks. Here, the security controls refer to the safeguards and protection capabilities appropriate for achieving particular security and privacy objectives. With the help of TARA, one can identify and prioritize possible threats against the target systems which can lead to security incidents. For detailed steps proposed to be taken by TARA please refer to [9].

There are connections between the analyses performed in TARA and HARA, which is the Hazard Analysis and Risk Assessment proposed for assuring functional safety of vehicles in ISO 26262-2:2018 E3.2 [6].

According to ISO DIS 21434, in a TARA the safety-relevant damage scenarios should be identified and selected, and the safety impact rating should be determined. In HARA the cybersecurity threats should be analyzed as hazard from a functional safety perspective in order to support the completeness of the hazard analyses and risk assessment and the safety goals. The TARA-HARA cross check process brings an alignment between the two analyses. They are triggered by cybersecurity experts collecting the damage scenarios after a first evaluation of the TARA that are assumed to be safety related. With safety experts and their consolidated HARA, they systematically crosscheck hazard events from the HARA against damage scenarios from the TARA. There are multiple options:

1) damage scenarios are comparable to some hazardous events. Then refinement of damage scenarios can come from a safety point of view with experience of catalogs of driving scenarios, and the damage scenarios can get refined through the systematic process of the HARA.
2) damage scenarios are just cybersecurity-oriented (no overlaps with the HARA). These damage scenarios are motivated by malicious attacks, not by faults of the SUT. This kind of end-consequences are not captured in HARA.

The design changes need to be applied and security controls have to be defined in early phases of development to avoid future rework. Security testing can be applied as part of the validation of these security controls and as part of the assessment of the final product against vulnerabilities.

Security testing in the automotive domain can be separated in four types depending on the phase of the product lifecycle:

**(1) Functional testing** is used for validation of the implementations of security controls.

**(2)** The TARA attack trees (with the goal as root node and ways of achieving that goal as paths from leaf nodes) provide a methodical way to describing the security of systems, based on varying attacks. The attack trees are used to create TARA attack paths, which define the interfaces for **protocol-guided automated or semi-automated fuzz testing**. The coverage of tested protocol can then be measured with percent.

**(3) Scanning and exploiting the implemented system security controls** for vulnerabilities helps to find out the system weaknesses. Subsequently, penetration tests will be conducted. The hardware, software or firmware penetration testing can exercise both physical and technical controls. Attack paths are identified including scanning, exploiting of unintended reproducible behaviors, and exploratory tests. The necessary level of testing is determined by the cybersecurity assurance level (CAL). In this paper, we focus on the black-box testing approach at concept level to test the impact of an attack on the overall system.

**(4) Asset-focused Red Team unbounded testing**, which is focused on one specific threat at a time and covers all kind of residual risks that were not addressed by controls or left open by implemented controls.

ISO/SAE DIS 21434 [4] recommends penetration testing as part of the cybersecurity validation requirements and recommendations [RC-10-03]. ISO/SAE DIS 21434 [4] also discusses penetration testing as a search method for vulnerabilities as part of product development integration and verification [RC-11-01].

*C. Hazard and Risk Analysis (HARA)*

In order to identify critical situations, we applied a Hazard and Risk Analysis as described in *ISO 26262* [6]. The identified functions are rated for the failure modes *No*, *Unintended*, *too Early*, *too Late*, *Less*, *More*, *Inverted* and *Intermittent*. The ratings of the assumed *Exposure (E)*, *Severity (S)* and *Controllability (C)* are used to compute the overall risk rating. The risk rating is then used to quantify the criticality of the risks based on the ASIL categorization defined in ISO 26262 [6]. These safety goals should be addressed in the derivation of attack descriptions. §III-B gives an example of a HARA.

III. SASEVAL APPROACH: SAFETY DRIVEN SECURITY VALIDATION

In this paper, we extend the existing notions of attacks by linking the validation to explicit safety goals. We introduce an attack derivation process that should identify for each target assets, potential threats and safety concerns. As the attacks are based on use cases formulated in natural language, their specification has to use natural language as well. We introduce the term *attack descriptions* to point out, that the attacks on this level do not include an implementation (either in a programming language nor a test script to be executed). The implementation will refine the attack description by incorporating the characteristics of SUT's interfaces and test stands, which are usually not determined yet within the concept phase into an executable representation. Our paper focuses on the derivation of attack descriptions and not on the actual implementation of these attacks.

An attack description has to consider potential operational modes of the system, as well as the operational context. Furthermore, the attack description may also anticipate a certain reaction/behavior of the SUT. This requires knowledge about the planned security controls of the architecture.

First of all, we describe how the derivation of an attack description is designed and how the completeness of the attack descriptions can be ensured.

The process for the derivation of an attack description has been set up to identify the information that needs to be provided to the four steps presented in Fig. 1:

1) **Threat Library Creation.** The threat library identifies threats that could be exploited in a certain scenario. By classifying threat scenarios according to threat types and then mapping these to different types of attacks (see §III-A), the library provides valuable inputs to the attack description process on the types of attacks that could be further detailed, described and implemented.

2) **Safety Concern Identification.** The safety concerns that need to be addressed by the validation as well as the safety artifacts used should be identified.

3) **Attack Description.** The safety concerns are used to select applicable attack types and corresponding threats of the threat library. The attack describes how the selected threat can be applied so that the safety goal associated with the safety concern could get violated.
   It has to be assured that in case of attacks, security controls react such that no safety goal is violated.

4) **Attack Implementation.** As the attack description is on the architectural level and has to be refined together with the subsequent development steps (i.e. the definition and specification of interfaces and behavior as well as the implementation). The actual implementation of the attack descriptions is out of scope of this paper.

Once the test objectives have been determined by the process steps 1 to 4, we can reason about the completeness of the malfunctions and critical situations. This deductive approach guarantees that the system is tested against critical unwanted effects. In order to complement the deductive approach by an inductive process, we propose to check whether all threats in the threat library are covered by the attack description. If an attack is not covered, the test engineer should consider either creating an additional attack description or writing a justification on why the threat is not applied for the given SUT. This inductive approach contributes to addressing all threats.

In the following subsections, we provide insights on each step and explain how the process is carried out in detail.

*A. Step 1 - Thread Library Creation*

In this section, we provide details about the creation of the threat library (Fig. 1), for which four sub-steps have been developed. Following the process, the different scenarios (§III-A1) are mapped to attack types (§III-A4), which facilitate the testing of these scenarios. An example of some automotive-related scenarios and their corresponding sub-scenarios is presented in Table I. These scenarios are used as a proof-of-concept to build a threat library.

The library could be useful especially in domains that share the same threat scenarios (§III-A2). This comes from the fact that the library facilitates mappings between the threat scenarios and threat types defined by the Microsoft STRIDE model [14] (§III-A3) as well as the ones between threat types and types of attacks (see §III-A4). The different stages of our process are now further detailed.

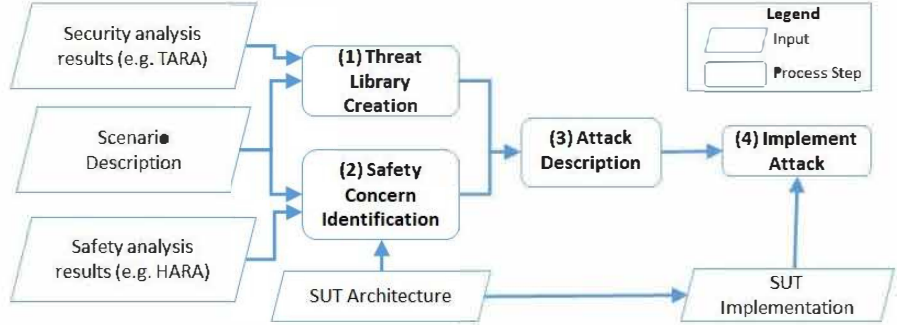

Fig. 1. Overview SaSeVAL approach: combine threat and safety driven process for attack derivation

TABLE I
EXAMPLE SCENARIOS CONNECTED TO THE AUTOMOTIVE DOMAIN.

| Scenario | Sub-scenario |
|---|---|
| Road intersection | An intersection with traffic lights is approached by a hijacked automated vehicle that has no intention to stop |
| | An automated vehicle approaches intersection which is equipped by a road-side system providing information about vulnerable road users. |
| | Emergency vehicle approaches a crowded intersection. |
| Keep car secure for the whole vehicle product lifetime | Vehicle updates are changes made to the hardware or software of a security, safety, or privacy relevant item that is deployed in the field. |
| Advanced access to vehicle | Demonstrator is reflecting the trend for property (vehicle) sharing. The traveler orders a car in the target destination via cloud-based service. |

TABLE II
SAMPLE ASSETS AND ASSET GROUPS FOR THE 3RD SCENARIO OF TABLE I

| Asset | Asset Groups |
|---|---|
| Gateway | Hardware |
| Driver and Maintenance personal | Person |
| ECU | Hardware/ Software |
| V2X communications | Information/ Hardware |

*1) Step 1.1 - Identification of useful scenarios:* In this step of the threat library creation process, we identify the general driving scenarios and sub scenarios that are considered to be the most useful for the development of the threat library. These scenarios are selected from the Scenario Description input illustrated in Fig. 1. The scenarios and sub scenarios could be compiled and detailed according to (i) Asset type, (ii) Asset groups, and (iii) Type of security analysis, if conducted (see Fig. 1) e.g. SAHARA [10], TARA [9] or HEAVENS [3].

Examples of assets are roadside unit database, in-vehicle signage system communication data, etc. In general, the number of assets for each scenario could be significant. However, assets could be classified into a set of asset groups having certain properties in common, which allows a simpler reference and classification of the concerns. Asset groups could be cloud services, devices, hardware, software, information, person, server, service, etc.. Certain assets could additionally be classified into asset types for the threat (see §III-A2).

The number of assets is substantial and several threat scenarios could be identified for each asset. Therefore, the assets and asset groups are used in the Stage 2 (see §III-A2) of the process to identify the candidate threat scenarios to focus on during Stage 3 (see §III-A3). Table II shows an example of some assets, the asset group they belong to.

*2) Step 1.2 - Identification of threat scenarios:* This step of the creation of the threat library consists of identification of different threat scenarios. These scenarios could be identified by studying the general scenarios, such as the ones listed in Table I alongside their assets. This could be done using threat analysis techniques such as TARA [9], HEAVENS [3] or SAHARA [10]. The techniques should ensure, that all threats are systematically identified (RQ1).

Table III shows examples of some threat scenarios identified after studying a specific scenario called "keep car secure for the whole vehicle product lifetime". It is worth noting that the number of threat scenarios that could be identified as a result of conducting a threat analysis such as SAHARA could be significantly large. Therefore, depending on the type of asset that is of interest, one could limit the list of threat scenarios and therefore contribute to the fulfillment of RQ2. For example, a certain asset could be classified into one of the following types and the threat analysis could be conducted only of those assets of interest:

- Generic assets: Assets relevant for multiple scenarios.
- Interesting asset from a certain use case's perspective. Examples of these use cases are presented in §IV.
- Generic for current vehicles: Assets which are generic for all current vehicles, thus having the highest priority.
- Generic for ADAS (Advanced driver-assistance systems)/AD (Autonomous driving): Assets which are generic for vehicles equipped with ADAS/AD systems.
- Generic for connected vehicles: Assets which are generic for vehicles that communicate bidirectionally with other systems outside of the vehicle.

TABLE III
THREAT SCENARIOS AND THREAT TYPES IDENTIFIED FOR THE "KEEP CAR
SECURE FOR THE WHOLE VEHICLE PRODUCT LIFETIME" SCENARIO.

| Threat Scenario | Threat Type (STRIDE) |
|---|---|
| Spoofing of messages by impersonation | Spoofing |
| External interfaces (such as USB) may be used as a point of attack, for example through code injection | Elevation of privilege |
| Manipulation of functions to operate systems remotely, such as remote key, immobiliser, and charging pile | Tampering |

TABLE IV
STRIDE THREATS AND ATTACKS.

| Threat Type | Attack Types |
|---|---|
| Spoofing | Fake messages, Spoofing |
| Tampering | Corrupt data or code, Deliver malware, Alter, Inject, Corrupt messages, manipulate, Config. change |
| Repudiation | Replay, Repudiation of message transmission, Delay |
| Information disclosure | Listen, Intercept, Eavesdropping, Illegal acquisition, Covert channel, Config. change |
| Denial of service | Disable, Denial of service, Jamming |
| Elevation of privilege | Illegal acquisition, Gain elevated access |

*3) Step 1.3 - Mapping of threat scenarios to threat types:* Threat scenarios could be directly mapped to attacks. However, these mappings could be done subjectively depending on how the scenarios are described. In order to perform the mapping systematically, in this step, we map the threat scenarios to threat types described in the Microsoft STRIDE model [14].

For each threat type, the corresponding manifestations of the threats, i.e. attack types, can be determined in the next step (see §III-A4). Table III shows examples of some threat scenarios identified after studying a specific scenario called "keep car secure for the whole vehicle product lifetime". The table also shows the mapping of the threat scenarios identified and the threat types following the STRIDE model.

*4) Step 1.4 - Mapping of threat scenarios to attack types according to the STRIDE model:* A mapping can be done from the STRIDE threat types to the corresponding manifestations of the threats, i.e. attack types. Table IV shows an overview of the mapping. Having conducted this mapping in this step finalizes the process of mapping the scenarios described in Fig. 1 to the potential attacks that could be used to evaluate the scenarios. The attack types could be a basis for the implementation of concrete attack vectors to evaluate the scenarios as done in §III-C. Hence, attack types provide the foundation for a detailed attack description resp. implementation. Therefore this step contributes to RQ3.

Table V shows an example of following the above four steps (see §III-A1 to III-A4) for mapping the assets of "keep car secure for the whole vehicle product lifetime" scenario to concrete attacks types. The table also shows high-level examples of the attacks that could be conducted.

*B. Step 2 - Safety Concern Identification*

The safety concern is determined via safety analysis. It expresses which kind of accident may happen, if it is not fulfilled. It serves as test objective that the validation should address. When SaSeVAL is applied on the vehicle level (as in our example) the safety goals are addressed. The safety goals are result of the HARA. The argumentation for achieving completeness (RQ1) is based on the guideword approach (see §II-C) for failure modes that has been applied.

The HARA is used to identify the hazards that the validation is supposed to address (RQ2). A higher ASIL rating may be used to justify a greater testing effort. Below is an excerpt of a HARA for a rating of hazard:

- **Function (with ID)** Hazardous location notifications (Road works warning) (Rat01)
- **Failure Mode and Hazard:** NO - The driver can not be warned and the automated control is not returned.
- **Exposure & Hazardous Event:** E=3 Crash into road works (see Statistics Road Works)
- **Severity:** S=3 Workers injured/dead
- **Controllability:** C=3 The driver is not supposed to monitor the road while automated driving mode is active
- **Safety Goal:** SG01. Avoid ineffective location notification without returning driving control to human (ASIL C)

It can be used to identify which situations may be critical for a specific failure mode, as well as to determine how the vehicle may react with appropriate security controls.

*C. Step 3 - Attack Descriptions*

An attack description operates on the concept level. Its purpose is to describe potential attacks and to proof during the tests, that the implemented security-controls prevent a violation of the safety goals (e.g. by completely mitigating the attack, closing identified attack paths or detect and attack and switch reliable to a safe state). Hence, the description has to name the safety goal as well as the threat scenario addressed.

In order to achieve precise and reproducible results (RQ3), we identified that the attack should contain the following information:

- **Attack description.** Description of the attack on concept level. It may include the attackers motivation and which goal the attack is pursuing.
- **Precondition.** The situation in which the attack can get started.
- **Expected Measures.** The measures may either be security controls in place to prevent the attack to be successful, or safety measure, which may act as a fallback if no security control is in place.
- **Attack Success.** The criteria for which the attack is successful. Depending on the attack it may an unavailability of the service, control over the SUT, or acceptance of manipulated data.
- **Attack Fails.** The tester should also specify how a failed attack can be detected. To name some examples the

TABLE V
SAMPLE MAPPING OF THREAT TYPES TO ATTACKS TYPES FOR ASSETS OF "KEEP CAR SECURE FOR THE WHOLE VEHICLE PRODUCT LIFETIME" SCENARIO.

| Asset | Threat Scenario | threat Type (STRIDE) | Attack type | Attack Examples |
|---|---|---|---|---|
| Gateway | Abuse of privileges by staff (insider attack) | Elevation of privilege | Gain elevated access | Technical staff creating backdoors or abusing their authorities. |
| Gateway | Code injection, e.g. tampered software binary might be injected into the communication stream | Tampering | Inject | Injection of communication data e.g. on the CAN communication link or corruption of payload. |
| ECU | External interfaces such as USB or other ports may be used as a point of attack, for example through code injection | Elevation of privilege | Gain unauthorized access | Connecting USB memories infected with malware to the infotainment unit. |
| ECU | Innocent victim (e.g. owner, operator or maintenance engineer) being tricked into taking an action to unintentionally load malware or enable an attack | Spoofing | Fake messages | Deceiving the user by sending an email pretending to be from the OEM, asking the user to download a malware and install it on the vehicle. |

SUT may ignore the malicious data or senders, create dedicated log files or degrade it's functionality.
- **Attack Implementation Comments.** The attack may contain additional information on how the attack could be implemented as soon as the architecture is refined to an implementation.

The test engineer would need to identify the interface of the asset given by the threat scenario that should be attacked. The attack and potential attack steps should be described so that an implementation is possible as soon as the interfaces are specified in detail.

The definition of preconditions determines the conditions in which the attack should be executed under. They are specified as conditions on the state of the environment or the operational mode of the vehicle. As the attack description is created before the system safety measures and security controls are finally set, it should be independent of the implementation.

For each such situation fulfilling the preconditions, the testing should be conducted using suitable stimuli to test whether the security controls are sufficient. For each combination of safety goal and attack type the potential attacks and the safety and/or security measures to be active are identified.

Each potential security control that facilitates achieving the test objective has to be identified. The preliminary architecture of the SUT is used as an input to determine which kinds of measure is present. The measures can be determined by safety or security analysis. For example, a safety measure could determine that plausibility checks fail and trigger the shutdown of a system. Such a measure could also be effective if an attack would cause inconsistent states. A security control could determine that an attack is occurring by monitoring for unintended or suspicious behaviour. Beyond reaching a safe state, the system may identify an attacking entity and classify it as not trustworthy for future collaboration. This could prevent other systems in the ecosystem from such attacks, making the overall system more resilient to attacks.

The assumed safety/security measures realized by the system are used to select suitable attack types and their corresponding threats from the threat library (see §III-A). In some cases, a test attacker only knows about potential attack vectors and how to potentially influence the system. In these cases, test engineers are supposed to identify safety/security measures that could be in place so that the attack can be designed to bypass these measures.

The test engineers may already anticipate potential measures, as they determine the postconditions for attack, which are specified as *attack success* and *attack fail* criteria. The two cases are differentiated, as the success case usually indicates how the safety goal is violated, while the failing case indicates a non-vulnerable system. That latter, however, should be further analyzed to see whether the system configuration or a dedicated protective measure prevents the attack from being successful and if it might by bypassed. The test's fail and pass criteria are determined for each test objective, taking the expected safety/security measures into account. Based on the test objective, the corresponding safety goal has one safe states that must be reached in case of malfunctions. Further, the safety goal's FTTI determines how fast the system should react to a given malfunction. The measure may also add information about the detection of attacks within the target SUT. For example, the measure may create a log-file if an attack is detected.

Additionally, a test engineer can add comments for the upcoming attack implementation. It may be stated how attacks are deployed, which and how communication networks could be attacked and how the test report is gathered.

*D. Attack Implementation*

The attack implementation could be conducted as soon as the implementation of the SUT is in place. It considers the SUT's interfaces as well as design decisions of the system's implementation, but is not within the scope of this paper.

IV. EVALUATION

The following two use cases have been worked out as part of the EU SECREDAS project [1] . For the complete use case description including ratings and attack descriptions please refer to Deliverable 3-10 [1] for "Demonstrator II" (here "Use Case I") and for "Keyless Car Opener" (here "Use Case II").

The terms use case and scenarios are used interchangeably within this paper.

*A. Use Case I - Autonomous Driving*

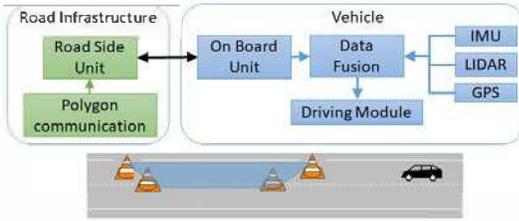

Fig. 2. Use Case I: Autonomous Vehicle approaches a construction side and returns control to the driver.

TABLE VI
ATTACK DESCRIPTION EXAMPLE FOR USE CASE I

| | |
|---:|---|
| Attack Description | AD20 - Attacker tries to overload the ECU by packet flooding. |
| SG IDs | SG01, SG02, SGO3 |
| Interface / ECU | OBU_RSU |
| Link to Threat Library | Threat scenario 2.1.4: An attacker alters the functioning of the Vehicle Gateway (so that it crashes, halts, stops or runs slowly), in order to disrupt the service |
| Types | Threat: Denial of Service - Attack: Disable |
| Precondition | Vehicle is approaching the construction side |
| Expected Measures | Message counter for broken messages |
| Attack Success | Shutdown of service |
| Attack Fails | Security control identifies unwanted sender enforce change of frequency |
| Attack impl. comments | Create an authenticated sender as attacker besides the original sender, additionally the attacker sender should send extra messages (with high frequency or in chaotic way) |

In this use case of we focus on a scenario where the autonomous vehicle approaches a construction site (see Fig. 2). The road side unit (RSU) informs the vehicle via the on board unit (OBU) about the upcoming site. The OBU should inform the driver, so that control is transferred back (upfront) to the driver. We analyzed potential attacks that may occur on the RSU-OBU interface.

We decided to conduct the analysis using HARA. An example line of the HARA conducted is given in §III-B. Within the HARA we identified a total of three functions: "Hazardous location notifications (Road works warning)", "Signage applications (In-vehicle speed limits)", "Warning of other traffic participants about hazardous vehicle state" leading to this safety goal. As failure modes may lead to more than one failure, we achieved in total 29 ratings for the example. The number of ratings is 5 for "N/A", 5 for "No ASIL", 7 for "ASIL A", 3 for "ASIL B", 7 for "ASIL C" and 2 for "ASIL D". Based on the application of the failure modes (see §II-C) and the ratings, we identified following safety goals:

- SG01. Avoid ineffective location notification without returning driving to the human (ASIL C),
- SG02. Avoid intermittent control switches (ASIL C),
- SG03. Communicate Speed Limits safely (ASIL D),
- SG04. Avoid missing take-over warnings (ASIL C),
- SG05. Avoid too many unintended warnings about hazardous vehicle states (ASIL B), and
- SG06. Avoid profile building with warnings (ASIL A)

These safety goals have been addressed in the derivation of attack descriptions. We identified for each combination of safety goal and attack type the potential attacks and which safety and/or security measure we expect to be active. Further, attack success, fail criteria and comments regarding a later implementation of the attacks have been identified.

The application of SaSeVAL yielded 23 attack descriptions. One example is given in Table VI. It shows that an denial of service attack could lead to the violation of safety goal SG01, SG02 and SG03, if an authenticated attacker would send extra data. The SUT is expected to detect the flooding situation and to react appropriately[1].

The application of the threat library also yielded attacks, which are not covered by classical security controls, such as encryption and authentication. For example, the category "Repudiation - Replay" could be addressed by an attack, in which warnings are replayed from other locations or other vehicles may lead to a violation of SG05.

The analysis and the reviews added were beneficial to to the project, as links between the attacks and potential safety impacts were made explicit. The systematic approach showed that a means for arguing completeness w.r.t. safety concerns is given. The integration plan considers the attack description as basis for testing as soon as the SUT and test stands is built.

*B. Use Case II - Keyless Car Opener*

TABLE VII
ATTACK DESCRIPTION EXAMPLE FOR USE CASE II

| | |
|---:|---|
| Attack Description | AD08 - The attacker uses modified keys to gain access to the vehicle. |
| SG ID and Name | SG01. Keep vehicle closed. |
| Interface / ECU | ECU_GW |
| Link to Threat Library | Threat scenario 3.1.4: Spoofing of messages (e.g. 802.11p V2X) by impersonation. |
| Types | Threat: Spoofing - Attack: Spoofing |
| Precondition | Vehicle is closed. Attacker has an authenticated communication link |
| Expected Measures | Check received vehicles electronic ID with list of allowed IDs |
| Attack Success | Open the vehicle |
| Attack Fails | Opening is rejected |
| Attack impl. comments | a) Randomly replace IDs of keys and b) test against increasing IDs (if a valid ID is known) |

For further evaluation, a keyless car opener has been analyzed. The use cases are opening and closing a vehicle via smartphone, which communicates via Bluetooth low energy with the car. Within the HARA, we analyzed these two functions. The rating has the same structure applied in the previous presented use case. The 20 ratings obtained yielded 7 N/A cases, 5 No-ASIL cases, 2 for ASIL A, 4 for ASIL B, 1 for ASIL C and 1 for ASIL D and resulted safety goals:

- SG01. Keep vehicle closed (ASIL D)
- SG02. Avoid intermittent open/close (ASIL B)
- SG03. Prevent non-availability of opening (ASIL A)
- SG04. Prevent unintended closing (ASIL A)

---
[1]Please notice, that cases like manipulated data or unauthenticated data are treated by other attack descriptions.

The application of the rating has shown that the approach is very well-suited to tackle safety concerns as identified in the HARA.

The attack descriptions are focused on attacks that may occur despite having a valid end-to-end encryption. This includes attacks like the exploitation of security vulnerabilities (for example in the Bluetooth stack), social engineering attacks, replay attacks and denial of service attacks. Table VII shows an exemplary attack description. The goal of the attack is to validate the key generation and validation mechanism to ensure an attackers can not generate valid keys on their own.

Other attacks identified specifics of the architecture, which are not covered by the encryption of the Bluetooth phone interface such as:

- Flooding of the CAN bus, by forwarded Bluetooth request, reducing availability of the function (SG03).
- Replaying of the opening command by an attacker (this might be prevented by timestamps resp. challenge-responds-patterns within the communication).

The application of SaSeVAL has identified in total 27 possible attacks with safety critical impact and additionally two attacks, which deal with privacy issues, as attacks may create profiles about the usage.

## V. CONCLUSION

The SaSeVAL approach systematically identifies test cases for the validation of safety critical system. It aims at identifying safety critical hazards, which might be introduced by security threats. Thereby, SaSeVAL especially addresses the needs of the automotive domain.

The approach assures that the security testing is complete, by identifying the threats and linking them to safety goal violations. The focus on safety allows us to define concrete attack descriptions which aim at maximizing the harm of a potential attack. In order to address privacy concerns, we propose to extend this work in the future.

The attack descriptions could be refined for actual test cases. As preparation for the refinement, we created a first version of a domain specific language (DSL). It encodes the attacks such that it can be automatically translated to test cases. The detailed description of the process address how research questions of completeness (RQ1), practical handling of the test space (RQ2) and definition of reproducible attacks (RQ3) are answered.

The scenarios investigated consider automotive specific concerns, such as the location of the vehicle, the encoding of location in messages, the characteristics of busses as limited bandwidth, different bandwidth capabilities of networks. The application on automotive systems requires to consider the characteristics beyond authentication and encryption of messages. While we focused here on the automotive domain, the SaSeVAL approach is generic and can be adapted for other safety critical domains.

Safety/security concerns in safety-critical systems should be addressed already in early development stages to integrate appropriate measures into the core concepts as part of the requirements and architecture. The early definition of attacks based on threat libraries and on risk analysis enables an informed discussion and design of required security controls and counter measures.